\documentclass[prb,onecolumn]{revtex4}
\usepackage{graphics}
\usepackage{picinpar}
\usepackage{epsfig}
\usepackage{bbm}
\usepackage{float}
\usepackage{amsmath}

\begin{document}
\title{Sample-to-sample fluctuations and bond chaos in the $m$-component spin glass}
\author{T. Aspelmeier}
 \affiliation{Max Planck Institute for Dynamics and Self Organization,
 Bunsenstr.\ 10, 37073 G\"ottingen, Germany}
 \affiliation{Scivis GmbH, Bertha-von-Suttner-Str.\ 5, 37085 G\"ottingen,
 Germany}
 \author{A. Braun}
\affiliation{Institut f\"ur Theoretische Physik, Universit\"at G\"ottingen,
Friedrich-Hund-Platz 1, 37077 G\"ottingen, Germany}

\begin{abstract}
We calculate the finite size scaling of the sample-to-sample fluctuations of the free energy $\Delta F$ of the $m$ component vector spin glass in the large-$m$ limit. This is accomplished using a variant of the interpolating Hamiltonian technique which is used to establish a connection between the free energy fluctuations and bond chaos. The calculation of bond chaos then shows that the scaling of the free energy fluctuations with system size $N$ is $\Delta F \sim N^\mu$ with $\frac{1}{5}\leq\mu <\frac{3}{10}$, and very likely $\mu=\frac 15$ exactly.
\end{abstract}

\maketitle

\section{Introduction}
Spin glass physics \cite{Mezard:1987} continue to be a very exciting and difficult topic. One of the important ongoing issues are the finite size corrections to thermodynamic quantities\cite{Parisi:1993,Parisi:1993a,Palassini:1999,Drossel:2000,Bouchaud:2003,Boettcher:2004a,Billoire:2006,Aspelmeier:2008c,Parisi:2009,Parisi:2010}. Such finite size corrections are usually impossible to calculate within replica theory (which is otherwise extremely successful for the spin glass and other problems), due to the massless modes which are often encountered and which prevent going beyond zero-loop order in a perturbation expansion. Nevertheless, many results are by now established for the Sherrington-Kirkpatrick spin glass\cite{Sherrington:1975}, either numerically or analytically (see references cited above).

An observable which is particularly interesting is the finite size
scaling of the sample-to-sample fluctuations of the free energy. This
quantity provides a link between two apparently distant fields, namely
spin glass physics and extreme value statistics\cite{Biroli:2007}. In
extreme value statistics, the question is the probability distribution
of extremal events such as the maximum (or minimum) of a set of random
numbers. The classic results of extreme value theory state that such
extremal events follow one of three possible limiting distributions
(the Weibull, Gumbel or Fr\'echet distribution). Recently, a fourth
universality class was found, the Tracy-Widom distribution for the
smallest (or largest) eigenvalue of a Gaussian random
matrix\cite{Tracy:2000}. Similarly, one could ask what the
distribution of the ground state energy of a (Sherrington-Kirkpatrick
or other) spin glass is, which is a question of extreme value
statistics. In a statistical mechanics setting, however, it can be
generalized to the question of what the distribution of the
\textit{free} energy is at finite temperature. This is a very
difficult question indeed (both for finite and zero temperature). To
keep it simpler, we merely ask what the \textit{width} of the
distribution is (i.e.\ the sample-to-sample fluctuations), and this
width $\Delta F$ will scale in some way with the system size $N$,
i.e.\ $\Delta F\sim N^\mu$ with an exponent $\mu$. These free energy
fluctuations have been considered for the Sherrington-Kirkpatrick
model by numerical investigations, see e.g.\
Refs.~\onlinecite{Bouchaud:2003,Boettcher:2004a,Parisi:2009}. There
also exist heuristic arguments for $\mu=\frac 14$ \cite{Bouchaud:2003}
and $\mu=\frac 16$\cite{Kondor:1983b,Crisanti:1992,Aspelmeier:2008c,Parisi:2010,Parisi:2010a},
and the limit $\mu\le\frac 14$ has been
shown\cite{Aspelmeier:2008,Aspelmeier:2008a}. All these results show
that the Sherrington-Kirkpatrick model does not fall in any of the
four established universality classes of extreme value statistics. For
a different famous replica symmetric spin glass model, the spherical
spin glass\cite{Kosterlitz:1976}, the situation is different. Since its
groundstate energy is the smallest eigenvalue of a Gaussian random
matrix, this model falls into the Tracy-Widom universality
class\cite{Andreanov:2004}. This implies that the fluctuations of the
free energy scale as $\Delta F\sim N^{1/3}$, which has been confirmed
recently by a replica calculation\cite{Parisi:2010}.

In this work we consider the fully connected $m$ component vector spin
glass in the limit of large $m$. This model is known to be replica
symmetric\cite{Almeida:1978b}, which suggests that its free energy fluctuations
might fall into the same universality class as the spherical
model. Furthermore it might be expected that due to replica symmetry the fluctuations in this model are
simpler to calculate than in the Ising spin glass. Unfortunately, this hope is not
entirely justified, as we will see. As mentioned above, it is usually
impossible to calculate subextensive quantities within
replica theory. We circumvent this problem by using a connection
between the sample-to-sample fluctuations of the free energy and bond
chaos, which was derived for the Ising spin
glass\cite{Aspelmeier:2008a}, and which we generalize here to vector
spin glasses. This connection allows us to calculate the
sample-to-sample fluctuations by calculating bond chaos instead, and
this is possible within a large deviation approximation, combining
techniques from Ref.~\onlinecite{Viana:1988} and
\onlinecite{Aspelmeier:2008b}. As it will turn out, the large
deviation approximation is not good enough to obtain the final answer,
and we must resort to additional resources, such as our previous
knowledge of the scaling properties of the large-$m$
model\cite{Hastings:2000,Aspelmeier:2004a,Lee:2005a,Braun:2006} and
some additional scaling assumptions. To forestall our main result, we
obtain $\mu\le\frac 15\le\frac{3}{10}$ and most likely $\mu=\frac 15$
exactly. This shows that the large-$m$	spin glass does not belong to any
of the four universality classes and probably to a different class than the Ising spin glass (unless it so happens that e.g.\ $\mu=\frac 14$ for both the Ising and the large-$m$ spin glass, which is not entirely ruled out by our results but which we deem unlikely). The result $\mu=\frac 15$ was first suggested in Ref.~\onlinecite{Hastings:2000}. However, the derivation required the existence of a gap in the eigenvalue spectrum of the inverse susceptibility matrix which was later shown not to exist\cite{Aspelmeier:2004a}. Here, we provide a different explanation for this value of $\mu$.

This paper is organized as follows. In section \ref{section_connection} we will show how to derive the exact connection between the sample-to-sample fluctuations of the free energy, $\Delta F_N$, and bond chaos for the $m$-component spin glass using interpolating Hamiltonians. We will see that the fluctuations only depend on the finite size scaling of averages of powers of the link overlap $q_L$ between two copies of a spin glass with different (but correlated) disorder. The averages of the link overlaps will be calculated in section \ref{calculating_P(q)} by calculating the probability distribution of the link overlap, $P_\epsilon(q_L)$, where the parameter $\epsilon$ measures the degree of correlation.  We will use these results in section \ref{fluctuations} to derive the exponent $\mu$ of the free energy fluctuations. 

\section{Connection between the sample-to-sample fluctuations and bond chaos}\label{section_connection}
We use the technique of interpolating Hamiltonians\cite{Guerra:2002} in order to derive an exact connection between the sample-to-sample-fluctuations of the free energy, $\Delta F_N$, and bond chaos in the $m$-component spin glass. To do so, we adapt the method from Ref.~\onlinecite{Aspelmeier:2008a}, which was developed for the Ising spin glass, to the $m$-component vector spin glass. The Hamiltonian of the $m$-component spin glass is
\begin{align}
\mathcal{H} &= \sqrt{\frac 1N} \sum_{i<j}J_{ij}\vec s_i \vec s_j,
\label{ham_basic}
\end{align}
where $\vec s_i$ are $m$-component spins and $J_{ij}$ are independent Gaussian random variables with unit variance. The spins are assumed to be normalized in such a way that $\vec s_i^2 = m$.

The main idea for calculating the sample-to-sample fluctuations of the free energy is to use the following two Hamiltonians,
\begin{align}
\mathcal{H}_t&=-\sqrt{\frac{1-t}{N}}\sum_{i<j}J_{ij}\vec s_i\vec s_j-\sqrt{\frac{t}{N}}\sum_{i<j}J_{ij}^{\prime}\vec s_i\vec s_j\hspace{2ex}{\rm and}\nonumber\\
\mathcal{H}_\tau^\prime&=-\sqrt{\frac{1-\tau}{N}}\sum_{i<j}J_{ij}\vec s_i\vec s_j-\sqrt{\frac{\tau}{N}}\sum_{i<j}J_{ij}^{\prime\prime}\vec s_i\vec s_j
,\label{ham}
\end{align}
with $0\leq t,\tau\leq 1$ and $J_{ij}^\prime$ and $J_{ij}^{\prime\prime}$ additional independent Gaussian random variables with unit variance, and express the free energy fluctuations in terms of them. These Hamiltonians interpolate between a spin glass with a given set of coupling constants $\{J_{ij}\}$ for $t=0$ (or $\tau=0$) and an identical spin glass with a different, independent set of coupling constants $\{J'_{ij}\}$ or $\{J''_{ij}\}$ for $t=1$ (resp.\ $\tau=1$). For all $0<t,\tau<1$ we have the same type of spin glass but with coupling constants $\sqrt{1-t}J_{ij}+\sqrt{t}J'_{ij}$ (or $\sqrt{1-\tau}J_{ij}+\sqrt{\tau}J''_{ij}$), which are still Gaussian random variables with unit variance. We denote the disorder average with respect to all coupling constants $\{J_{ij}\}$, $\{J'_{ij}\}$ and $\{J''_{ij}\}$ as $E\cdots$. Thermal averages will be denoted as $\langle\cdots\rangle$.

The free energy fluctuations can be written as $\beta^2\Delta F_N^2=E(\log Z_1-\log Z_0)(\log Z_1^\prime-\log Z_0^\prime)$, with $Z_t$ and $Z_\tau^\prime$ being the partition functions with respect to the Hamiltonians $\mathcal{H}_t$ and $\mathcal{H}_\tau^\prime$. This can be seen by writing $\log Z=-\beta F$ and using the independence of the different sets of bonds for $E F_1 F'_0 = E F_0 F'_1 = E F_1 F'_1 = \overline{F}^2$ and $E F_0 F'_0 = \overline{F^2}$ (the overbar denotes a disorder averaged quantity of the original system of interest, i.e.\ Eq.~\eqref{ham_basic}). We use the idea in Ref.~\onlinecite{Guerra:2002} to represent this expression by integrals in the form
\begin{align}
\beta^2\Delta F_N^2&=\int_0^1dt\int_0^1d\tau E\frac{\partial}{\partial t}\log Z_t\frac{\partial}{\partial\tau}\log Z_\tau^\prime.\label{connection}
\end{align}
To calculate the right hand side, we follow Ref.~\onlinecite{Aspelmeier:2008a} and obtain
\begin{align}
E\frac{\partial}{\partial t}\log Z_t\frac{\partial}{\partial\tau}\log Z_\tau^\prime&=\frac{\beta^2}{16N^2}
 E\sum_{ijkl}\big(\langle(\vec s_i \vec s_j)(\vec s_k \vec s_l)\rangle_t-\langle\vec s_i \vec s_j\rangle_t\langle\vec s_k \vec s_l\rangle_t\big)\big(\langle(\vec s_i \vec s_j)(\vec s_k \vec s_l)\rangle_\tau-\langle\vec s_i \vec s_j\rangle_\tau\langle\vec s_k \vec s_l\rangle_\tau\big)\nonumber\\
&+\frac{\beta^2}{8N\sqrt{1-t}\sqrt{1-\tau}}\Big(E\sum_{ij}\big(\langle\vec s_i\vec s_j\rangle_t\langle\vec s_i\vec s_j\rangle_\tau\big)-m^2N\Big)\label{appendix}.
\end{align}
The subscript $t$ or $\tau$ on the thermal averages indicates whether the average is to be taken in a system with Hamiltonian $\mathcal H_t$ or $\mathcal H'_\tau$. 

We introduce the link overlap between two replicas with potentially different coupling constants as $q^{13}_L=\frac{2}{N(N-1)}\sum_{i<j}\langle\vec s_i\vec s_j\rangle_t\langle\vec s_i \vec
s_j\rangle_\tau=\frac{1}{N(N-1)}\Big(\sum_{\mu\nu}\langle\big(\sum_i s_{i\mu}^{(1)} s_{i\nu}^{(3)}\big)^2\rangle-m^2N\Big)$. As we will shortly need not only $2$ but up to $4$ replicas, we label the replicas with upper indices $(1)$ to $(4)$, where replicas $1$ and $2$ have Hamiltonian $\mathcal H_t$ and replicas $3$ and $4$ have Hamiltonian $\mathcal H'_\tau$. Analogously to $q^{13}_L$, we define the other possible link overlaps $q^{14}_L$, $q^{23}_L$ and $q^{24}_L$. Then equation (\ref{appendix}) can be expressed in the following way
\begin{align}
E\frac{\partial}{\partial t}\log Z_t\frac{\partial}{\partial\tau}\log Z_\tau^\prime&=
\frac{(N-1)^2\beta^4}{16}E\langle (q_L^{13}-q^{23}_L)(q_L^{13}-q_L^{14})\rangle+\frac{(N-1)\beta^2 }{8\sqrt{1-t}\sqrt{1-\tau}}E\langle q_L^{13}\rangle\nonumber\\
&=\frac{(N-1)^2\beta^4}{16}\Big([ (q_L^{13})^2 ] - [ q^{13}_L] ^2\Big)+\frac{(N-1)\beta^2 }{8\sqrt{1-t}\sqrt{1-\tau}}[ q^{13}_L].
\label{preparing_connection}
\end{align}
The notation $[\dots]$ in the last line stands for the average taken with the bond averaged probability distribution $P_\epsilon(q_L)$ of finding a given link overlap $q_L$. The parameter $\epsilon$ indicates the statistical ``distance'' between the sets of bonds in the two replicas involved and will be defined in detail below. In principle, one would have to consider simultaneous multi-replica overlaps (such as, for instance, the term $E\langle q_L^{13} q_L^{23}\rangle$). Fortunately, it follows  from replica symmetry, which holds for the $m$-component spin glass in the large-$m$ limit, that the corresponding joint probability distribution $P^{123}_\epsilon(q_L^{13},q_L^{23})$ factorizes into $P^{123}_\epsilon(q_L^{13},q_L^{23}) = P_\epsilon(q_L^{13})P_\epsilon(q_L^{23})$. A similar statement holds for the four-replica probability distribution $P^{1234}_\epsilon(q_L^{13},q_L^{24})$. 

Up to now the overlaps $q^{ab}_L$ depend on the two parameters $t$ and $\tau$. However, the only important quantity is how much the sets of bonds in the two replicas $a$ and $b$ differ. For example, when $t=\tau=0$, the bonds in replicas 1 and 3 (say) are identical. On the other hand, when $t=0$ and $\tau=1$ (or vice versa), the bonds in replicas 1 and 3 are completely uncorrelated. The degree of correlation between the two sets of bonds can be measured by the one parameter $\epsilon$ defined by $\frac{1}{\sqrt{1+\epsilon^2}}=\sqrt{1-t}\sqrt{1-\tau}$, as shown in Ref.~\onlinecite{Aspelmeier:2008a}. When $\epsilon=0$, the bonds are identical; when $\epsilon=\infty$, the bonds are completely uncorrelated. 

The parameter $\epsilon$ can be used to eliminate the integration variable $\tau$ from Eq.~\eqref{connection} by a variable transformation.  The integral over $t$ can then be evaluated exactly, and as a result, the connection between the sample-to-sample-fluctuations of the free energy and bond chaos in the $m$-component spin glass is found to be (see Ref.~\onlinecite{Aspelmeier:2008a} for details)
\begin{align}
\beta^2\Delta F_N^2&=\frac{(N-1)^2\beta^4}{16}\int_0^\infty d\epsilon f_2(\epsilon)\big([ (q_L^{13})^2] -[ q_L^{13}]^2\big)+\frac{(N-1)\beta^2}{4}\int_0^\infty d\epsilon g_2(\epsilon)[q_L^{13}],\nonumber\\
&=: \frac{(N-1)^2\beta^4}{16} I_{21} + \frac{(N-1)\beta^2}{4} I_{22}
\label{final_connection}
\end{align}
with the nonnegative functions $f_2(\epsilon)=\frac{2\epsilon\log(1+\epsilon^2)}{(1+\epsilon^2)^2}$ and $g_2(\epsilon)=\frac{\epsilon\log(1+\epsilon^2)}{(1+\epsilon^2)^{3/2}}$. Note that bond chaos enters Eq.~(\ref{final_connection}) through the measure of distance $\epsilon$ of the bonds of the two replicas between which the link overlap $q_L^{13}$ is calculated. The integrals $I_{21}$ and $I_{22}$ will be calculated below.

The analog of Eq.~\eqref{final_connection} was called ``second route to chaos'' in Ref.~\onlinecite{Aspelmeier:2008a}, hence the first index is $2$ on $I_{21}$ and $I_{22}$. In addition to this result, however, there is a another exact relation between the fluctuations and bond chaos (the ``first route to chaos''). It stems from using only the first interpolating Hamiltonian of equation (\ref{ham}), $\mathcal{H}_t$, and the relation $2\beta^2\Delta F_N^2 = E(\log Z_1 - \log Z_0)^2$, which is easy to derive. Proceeding similarly to above, one can prove the following equality:
\begin{align}
\beta^2\Delta F_N^2 &= -\frac{(N-1)^2\beta^4}{16}\int_0^\infty d\epsilon f_1(\epsilon)\big([ (q_L^{13})^2] -[ q_L^{13}]^2\big)+\frac{(N-1)\beta^2}{4}\int_0^\infty d\epsilon g_1(\epsilon)[q_L^{13}]\nonumber\\
&=: -\frac{(N-1)^2\beta^4}{16} I_{11} + \frac{(N-1)\beta^2}{4} I_{12}
\label{final_connection2}.
\end{align}
The only difference between this equation and Eq.~\eqref{final_connection} is the minus sign in front of the first term and the weight functions $f_1(\epsilon)=\frac{4\epsilon^2}{(1+\epsilon^2)^2}\arcsin\frac{1}{\sqrt{1+\epsilon^2}}$ and $g_1(\epsilon)=\frac{2}{(1+\epsilon^2)^{3/2}}\arcsin\frac{1}{\sqrt{1+\epsilon^2}}$ in the integrals $I_{11}$ and $I_{12}$ instead of $f_2(\epsilon)$ and $g_2(\epsilon)$ as in $I_{21}$ and $I_{22}$. The minus sign of the first term implies that the second term is an upper bound of $\beta^2\Delta F_N^2$. 

\section{Calculating $P_\epsilon(q_L)$}\label{calculating_P(q)}

Eq.~\eqref{final_connection} involves moments of the link overlap taken with the probability density $P_\epsilon(q_L)$. This function can in principle be calculated by taking two replicas with bond realizations drawn with parameter $\epsilon$ and constraining the replicas to have link overlap $q_L$. This constrained system has free energy $F_{\epsilon,J}(q_L)$, and the (non disorder averaged) probability density $P_{\epsilon,J}(q_L)$ then follows to be
\begin{align}
P_{\epsilon,J}(q_L) &= \frac{\exp(-\beta F_{\epsilon,J}(q_L))}{\int_0^\infty dq'_L\exp(-\beta F_{\epsilon,J}(q'_L))}.
\end{align}
Finally, $P_{\epsilon,J}(q_L)$ must be averaged over the disorder.

Unfortunately, this task is too difficult in general. Instead, we will calculate only the disorder averaged extensive part of the free energy, denoted by $Nmf_\epsilon(q_L)$, and the probability density defined by
\begin{align}
P^0_\epsilon(q_L) &= \frac{\exp(-\beta Nmf_\epsilon(q_L))}{\int_0^\infty dq'_L \exp(-\beta Nmf_\epsilon(q'_L))}.
\end{align}
This is the large deviation approximation to $P_\epsilon(q_L)$. Averages taken with respect to $P_\epsilon^0(q_L)$ will be denoted by $[\dots]_0$. The tail of this distribution will be the same as that of $P_\epsilon(q_L)$, but in general there will be deviations. In fact, this is the point where this paper differs most from Ref.~\onlinecite{Aspelmeier:2008a} since in that publication, the difference between the true and the approximative distribution was only quantitative, whereas here it is substantial. We know that for $\epsilon=0$ the link overlap distribution consists of a $\delta$ peak at the Edwards-Anderson value and $0$ elsewhere since the large-$m$ spinglass is replica symmetric. We will see below, however, that we do not observe this peak in $P^0_\epsilon(q_L)$ at all. It must therefore be generated from the finite size corrections to the extensive part of the free energy. 

Hence the finite size corrections are very important in this calculation, but we have no direct way of calculating them. In order to overcome this problem, we will show that $P_\epsilon^0(q_L)$ will be valid for $\epsilon$ larger than some crossover value, and we will use additional arguments and simulation results to fill the gap for smaller $\epsilon$.

\subsection{Replica calculation}
The first task is to calculate the extensive part of the disorder averaged free energy
$Nmf_\epsilon(q_L)$ of two replicas constrained to have link overlap $q_L$. To this end we calculate the partition function
$Z_{\epsilon,J}(q_L)$ of this two replica system. This system has vector spins $\vec s_i^x$ with $(\vec s_{i}^x)^2=m$ for every spin $i=1,\ldots ,N$ and the two real replicas
$x=(10),(2\epsilon)$ (this notation denotes the replica number in its first
entry and the value of $\epsilon$ in its second entry). These two
replicas differ in their coupling constants
$K_{ij}(\xi)=\frac{1}{\sqrt{1+\xi^2}}K_{ij}^0+\frac{\xi}{\sqrt{1+\xi^2}}K_{ij}^\prime$, where $K_{ij}^0$ and $K'_{ij}$ are independent Gaussian random numbers with unit variance,
by choosing $\xi=0$ for the first replica and 
$\xi=\epsilon$ for the second replica. Furthermore, the two replicas have link overlap $q_L$, which is enforced by a $\delta$ function in the partition function as follows:
\begin{multline}
Z_{\epsilon,J}(q_L) = {\rm Tr}_{\vec s}\Big(\delta\big(q_L-\frac{1}{N(N-1)}\sum_{\mu\nu}\big(\sum_i s_{i\mu}^{(10)} s_{i\nu}^{(2\epsilon)}\big)^2-\frac{m^2}{N-1}\big)\Big)\\ 
\times\exp\Big(\frac{\beta}{\sqrt{N}}\sum_{i<j}K_{ij}(0)\vec s_i^{(10)}\vec s_j^{(10)}+\frac{\beta}{\sqrt{N}}\sum_{i<j}K_{ij}(\epsilon)\vec s_i^{(2\epsilon)}\vec s_j^{(2\epsilon)}\Big)
\end{multline}
We follow the replica calculation of Viana\cite{Viana:1988} for the $n$ times replicated partition function, write the $\delta$-function in an integral representation with parameters $z_\alpha$, and introduce the traceless tensor $T_\alpha^{\mu\nu}$\cite{Viana:1988} to separate the diagonal part of $Q_{\alpha\beta}^{\mu\nu}$ (with $\sum_{\alpha,\beta\atop\mu,\nu}Q_{\alpha\beta}^{\mu\nu}=2\sum_{\alpha < \beta\atop\mu ,\nu}Q_{\alpha\beta}^{\mu\nu}+2\sum_{\alpha ,\mu <\nu}T_\alpha^{\mu\nu}+\sum_{\alpha,\mu}\big(Q_{\alpha\alpha}+T_{\alpha}^{\mu\mu}\big)$). After evaluating the trace and regarding only terms up to third order in the tensors $Q$, $T$ and $R$ we get
\begin{align}
EZ_{\epsilon,J}^n(q_L)&\propto
\int\big(\prod_{\alpha} dz_\alpha\big)e^{-N\sum_{\alpha}q_L
  z_{\alpha}-m^2\sum_\alpha z_\alpha} \int d\Lambda_{\alpha\beta}^{\mu\nu}\exp\Big(N\Big[\tau\sum_{\alpha <\beta \atop \mu,\nu}\big(Q_{\alpha\beta}^{\mu\nu }\big)^2+\big(\tau+\frac{r}{m+2}\big)\sum_{\alpha, \mu <\nu}\big(T_{\alpha}^{\mu\nu }\big)^2\nonumber\\
&+\big(\tau+\frac{r^\prime}{m+2}\big)\sum_{\alpha, \mu}\big(T_{\alpha}^{\mu\mu }\big)^2
+\tau^\prime\sum_{\alpha\neq\beta\atop\mu,\nu}\big(R_{\alpha\beta}^{\mu\nu}\big)^2+\sum_{\alpha,\mu,\nu}\big(\frac{2z_\alpha}{\beta^2}+\frac{1}{\sqrt{1+\epsilon^2}}\big)^{(-1)}\big(R_{\alpha\alpha}^{\mu\nu}\big)^2\nonumber\\
&+\frac{\omega}{3!}\Big(\sum_{\alpha <\beta <\gamma \atop \mu,\nu,\rho}Q_{\alpha\beta}^{\mu\nu }Q_{\beta\gamma}^{\nu\rho }Q_{\gamma\alpha}^{\rho\mu }+\frac{ m^2}{(m+2)(m+4)}\sum_{\alpha\atop \mu <\nu <\rho}T_{\alpha}^{\mu\nu }T_{\alpha}^{\nu\rho }T_{\alpha}^{\rho\mu }+\frac{15 m^2}{(m+2)(m+4)}\sum_{\alpha,\mu}\big(T_{\alpha}^{\mu\mu}\big)^3\nonumber\\
&+\frac{3 m}{m+2}\sum_{\alpha <\beta\atop \nu,\mu <\rho}Q_{\alpha\beta}^{\mu\nu }Q_{\beta\alpha}^{\nu\rho}T_{\alpha}^{\rho\mu}+3 \sum_{(\alpha <\beta )\neq\gamma\atop \mu,\nu,\rho}Q_{\alpha\beta}^{\mu\nu }R_{\beta\gamma}^{\nu\rho}R_{\gamma\alpha}^{\rho\mu}+3 \sum_{\alpha <\beta\atop \mu,\nu,\eta}Q_{\alpha\beta}^{\mu\nu }R_{\beta\alpha}^{\nu\eta}R_{\alpha\alpha}^{\eta\mu}\nonumber\\
&+\frac{3 m}{m+2} \sum_{\alpha\neq\beta\atop (\mu <\nu),\rho}T_{\alpha}^{\mu\nu }R_{\alpha\beta}^{\nu\rho}R_{\beta\alpha}^{\rho\mu}+3 \sum_{\alpha\atop \mu<\nu,\eta}T_{\alpha}^{\mu\nu }R_{\alpha\alpha}^{\nu\eta}R_{\alpha\alpha}^{\eta\mu}
\Big)\Big]\Big),\label{final}
\end{align}
with $\tau=\frac{1}{2}(1-\frac{1}{\beta^2})$, $\tau^\prime=\frac{1}{2}(1-\frac{\sqrt{1+\epsilon^2}}{\beta^2})$, $r=-1$, $r^\prime=2m-1$, $\omega=1$ and the matrix $\Lambda_{\alpha\beta}^{\mu\nu}$ is given by
\begin{align}
\Lambda&=\begin{pmatrix} Q^{(10)} & R \\ R & Q^{(2\epsilon)}\end{pmatrix},
\end{align}  
with the above separation for $Q_{\alpha\beta}^{\mu\nu}$ and only
$T_\alpha^{\mu\mu}$ on the diagonal of the matrices $Q$. We split the
tensor $R$ into its diagonal matrices
$p_d^{\mu\nu}=R_{\alpha\alpha}^{\mu\nu}$ and the rest. This equation is solved by a saddle point integration for the tensors $Q_{\alpha\beta}^{\mu\nu}$, $T_{\alpha}^{\mu\nu}$ and $R_{\alpha\beta}^{\mu\nu}$ and the parameters $z_\alpha$.
\subsection{Solving the saddle point equations}
Solving the saddle point equations is the remaining task to derive the characteristic form of the overlap distribution $P_\epsilon(q_L)$. As the $m$-component spin glass was shown to be replica symmetric in the limit $m\to\infty$\cite{Almeida:1978b}, we calculate the saddle points in the replica symmetric case. The ansatz for a replica symmetric scenario is as follows \cite{Viana:1988,Aspelmeier:2008b}:
\begin{align}
Q_{\alpha\beta}^{\mu\nu}&=Q\delta_{\mu\nu},\hspace{2ex}T_{\alpha}^{\mu\nu}=0\nonumber\\
R_{\alpha\beta}^{\mu\nu}&=P\delta_{\mu\nu},\hspace{2ex}p_{d}^{\mu\nu}=p_d\delta_{\mu\nu}\hspace{2ex}{\rm
and}\hspace{2ex}z_\alpha=z
\end{align}
To briefly justify these equations, one has the usual interpretation $Q_{\alpha\beta}^{\mu\nu (10)}=\langle s_{\alpha\mu}^{(10)}s_{\beta\nu}^{(10)}\rangle$ from the replica calculation. Due to the isotropy of the model, averaged quantities like this reduce to $x\delta_{\mu\nu}$ with some mean value $x$, depending on the quantity at hand. Thus $EZ^n_{\epsilon,J}(q)$ is given by
\begin{align}
EZ_{\epsilon,J}^n(q_L)&\propto\int \big(\frac{N-1}{2\pi}\big)^ndz e^{-2Nnq_Lz-m^2nz}\int d\Lambda_{\alpha\beta}^{\mu\nu}\exp\Big(N\Big[\tau mn(n-1)Q^2+\tau^\prime nm(n-1)P^2+\frac{nm}{2} p_d^2\nonumber\\
&-mn\frac{1}{2\beta^2}\big(\frac{2z}{\beta^2}+\frac{1}{\sqrt{1+\epsilon^2}}\big)^{(-1)}p_d^2+2\frac{\omega n(n-1)(n-2)}{3!}mQ^3\nonumber\\
&+\omega n(n-1)(n-2)mQP^2+2\omega n(n-1)mQPp_d\Big]\Big)\label{Z_final}
\end{align}
This leads to four different saddle point equations (we use $Nm$ as the large parameter)
\begin{align}
0&=\tau Q- Q^2- P^2+ Pp_d\label{saddle_Q}\\
0&=\tau^\prime P-2QP+ Qp_d\label{saddle_P}\\
0&=p_d-\frac{1}{\beta^2}p_d\big(\frac{2z}{\beta^2}+\frac{1}{\sqrt{1+\epsilon^2}}\big)^{(-1)}-2 QP\\
\frac{q_L}{m}&=\Big(\frac{p_d}{\beta^2\big(\frac{2z}{\beta^2}+\frac{1}{\sqrt{1+\epsilon^2}}\big)}\Big)^2,
\end{align}
where line 3 and line 4 can be combined to
\begin{align}
p_d-2QP=\sqrt{\frac{q_L}{m}}\label{saddle_p_d}.
\end{align}
In the following, we substitute $\sqrt{\frac{q_L}{m}}\rightarrow q$.
 
From equation (\ref{Z_final}) the free energy is (after taking the usual replica limit $n\to 0$)
\begin{align}
\beta f_\epsilon(q)&={\rm const.}+p_dq-\frac{p_d^2}{2}-\frac{q^2}{2(1-2\tau^\prime)}+\tau Q^2+\tau^\prime P^2-\frac{2}{3}Q^3-2QP^2+2QPp_d\label{free_energy}
\end{align}
\subsubsection{Above and at the critical temperature}
Above ($\tau < 0$) and at the critical temperature ($\tau=0$), the Ising spin glass is replica symmetric, just as the $m$-component spin glass. Therefore the solutions of the saddle point equations are the same as in Ref.~\onlinecite{Aspelmeier:2008b}. Inserting them into the free energy (equation (\ref{free_energy})) gives
\begin{align}
{\rm above\hspace{1ex} T_c:\hspace{2ex} }\beta f_\epsilon(q)&=\frac{q^2}{2}\Big(1-\frac{\beta^2}{\sqrt{1+\epsilon^2}}\Big)+\mathcal{O}(q^4)\\
{\rm at\hspace{1ex} T_c:\hspace{2ex}  }\beta f_\epsilon(q)&=\Big\{{\frac{1}{6}q^3+\frac{3\epsilon^2}{16}q^2+\mathcal{O}(\epsilon^4,q^4)\hspace{4ex}\epsilon^2\ll q\ll 1\atop\frac{q^2}{2}\Big(1-\frac{1}{\sqrt{1+\epsilon^2}}\Big)+\mathcal{O}(q^4)\hspace{4ex}q\ll \epsilon^2\ll 1}
\end{align}
At $T_c$ the probability distribution $P_\epsilon (q)\propto e^{-Nm\beta f_\epsilon (q)}$ consists of two parts with different dominating exponents depending on $\epsilon$ being smaller or larger than $N^{-1/6}$:
\begin{align}
P_\epsilon (q)&\propto\begin{cases}e^{-Nm\frac{q^3}{6}}& \epsilon\ll N^{-1/6}\\ e^{-Nm\frac{q^2\epsilon^2}{4}}&N^{-1/6}\ll\epsilon\end{cases}
\end{align}  
\subsubsection{Below the critical temperature}
The equations (\ref{saddle_Q}), (\ref{saddle_P}) and (\ref{saddle_p_d}) can not be solved for general $q$ and $\epsilon$. Due to this, we calculate corrections to the two following solvable cases, $q=0$ and $\epsilon=0$, perturbatively in various limits.
The exact result for $q=0$ is
\begin{align}
P&=0\nonumber\\
Q&=\tau\label{q=0}\\
p_d&=0\nonumber
\end{align}\\
To find the solution for $\epsilon =0$, we rewrite equations (\ref{saddle_Q}) and (\ref{saddle_P}) in terms of the new variables $a=Q+P$ and $b=Q-P$
\begin{align}
(\tau+p_d)a-\frac{\tau-\tau^\prime}{2}(a-b)-a^2&=0\label{saddle_a}\\
(\tau-p_d)b-\frac{\tau-\tau^\prime}{2}(a-b)-b^2&=0\label{saddle_b}
\end{align}
with the solution ($\epsilon=0$, so $\tau=\tau^\prime$)
\begin{align}
Q&=\tau\\
P&=p_d=\frac{q}{1-2\tau},
\end{align}
for $a\neq 0$ and $b\neq 0$. With equation (\ref{free_energy}) we get $\beta f_0(q)=\beta f_0=\frac{1}{3}\tau^3$. We will use this free energy as the reference free energy as it is the energy of the unconstrained system. 
For $q>\tau-2\tau^2=q_{EA}$ the solution with $b=0$ maximizes the free energy (the solution with $a=0$ is an unphysical one). In terms of $\Delta q=q-(\tau-2\tau^2)$ it is (to lowest order in $\Delta q$)
\begin{align}
Q&=P=\frac{\tau+p_d}{2}\nonumber\\
p_d&=\tau+\frac{\Delta q}{1-2\tau},
\end{align}
and the difference to $\beta f_0$ is
\begin{align}
\Delta f_0(q)&=c_0\Delta q^3,
\label{beyondEA}
\end{align}
with $c_0=\frac{1}{6(1-2\tau)^3}$. Now we will calculate corrections to the first solution, $f_0$, in various different limits of $\epsilon$ and $q$.\\
\subsubsection{$\epsilon\to\infty$}
In the limit $\epsilon\to\infty$, i.e. $\tau^\prime\to -\infty$ Eq.~(\ref{saddle_P}) yields the solution (to leading order)
\begin{align}
P&=-\frac{Qp_d}{\tau^\prime}+\mathcal{O}(\frac{1}{(\tau^\prime)^2})
\end{align} 
For equation (\ref{saddle_Q}) this leads to (again in leading order)
\begin{align}
0&=(\tau+\frac{p_d^2}{\tau^\prime})Q-Q^2
\end{align}
For $\tau^\prime=-\infty$, $p_d$ is equal to $q$ and the free energy difference to $f_0$ is
\begin{align}
\beta f_\infty(q)-\beta f_0&=qp_d-\frac{p_d^2}{2}+\mathcal{O}(p_d^4)=\frac{q^2}{2}+\mathcal{O}(q^4)
\end{align}
\subsubsection{Perturbative solution for $\epsilon^2\ll q\ll 1$}
We use the saddle point equations for $a$ (\ref{saddle_a}) and $b$ (\ref{saddle_b}), insert $a=a_0+a_1$, $b=b_0+b_1$ and $p_d=p_{d0}+p_{d1}$, where the index $0$ denotes the undisturbed solution ($\epsilon=0$) and the index $1$ the first order corrections to it. The results are ($\Delta\tau=\tau-\tau^\prime > 0$)
\begin{align} 
{a_1=-\frac{\Delta\tau p_{d0}-p_{d1}a_0}{\tau+p_{d0}}\atop b_1=\frac{\Delta\tau p_{d0}-p_{d1}b_0}{\tau-p_{d0}}}\Rightarrow&{Q_1=\frac{\epsilon^2 q^2}{4\tau^2(1-2\tau)}\atop P_1=-\frac{\epsilon^2 q}{4\tau}} 
\end{align}
and the correction to $p_d$ is $p_{d1}=-\frac{\epsilon^2q}{2}$. For the free energy these results yield:
\begin{align}
\beta f_\epsilon(q)-\beta f_0&=\frac{2\tau +1}{16\tau}\epsilon^4 q^2+\mathcal{O}(\epsilon^4q^4)\label{f_1}
\end{align}
\subsubsection{Perturbative solution for $q\ll \epsilon^2\ll  1$}
In this case the suitable reference solution is the one with $q=0$, equations (\ref{q=0}). We again introduce corrections ($Q\to Q+\Delta Q$) to this solution and combine equations (\ref{saddle_P}) and (\ref{saddle_p_d}) in lowest order to
\begin{align}
\tau^\prime\Delta P-2Q\Delta P+Q(\Delta q+2Q\Delta P)&=0,
\end{align}
where $\Delta q$ is understood to be the correction to $q=0$. The solution is
\begin{align}
\Delta P=-\frac{\tau\Delta q}{\tau^\prime-2\tau+2\tau^2}&\stackrel{\epsilon^2\ll 1}{\to}\frac{\Delta q}{1-2\tau}-\frac{\epsilon^2\Delta q}{4\tau(1-2\tau)},
\end{align}
$\Delta Q=0$ and
\begin{align}
\Delta p_d=\Delta q\frac{\tau^\prime-2\tau}{\tau^\prime-2\tau+2\tau^2}&\stackrel{\epsilon^2\ll 1}{\to}\frac{\Delta q}{1-2\tau}-\frac{\epsilon^2\Delta q}{2(1-2\tau)}.
\end{align}
The result for the free energy then is
\begin{align}
\beta f_\epsilon(q)-\beta f_0&=\frac{1}{16\tau(1-2\tau)}\epsilon^4 q^2+\mathcal{O}(q^3).\label{f_2}
\end{align}
The restriction $\epsilon^2\ll 1$ is basically unnecessary and we could calculate the free energy in the limit $q\ll \min (1,\epsilon^2)$. However, since the regime of large $\epsilon$ will not contribute to the sample-to-sample-fluctuations we neglect it already at this point, and write $\beta f_\epsilon(q)-\beta f_0(q)=f(\epsilon)q^2$ with $f(\epsilon)=\epsilon^4 \frac{1}{16\tau(1-2\tau)}$ for small $\epsilon$.

\subsection{$P^0_\epsilon(q)$}
Although we are interested in the probability density of the link overlap $q_L$, the more practical quantity for the calculation turned out to be $q=\sqrt{q_L/m}$. We therefore formulate our results in terms of $q$ instead of $q_L$ for the time being. This is not a serious restriction since we have the simple relation
\begin{align}
[(q_L^{13})^n]_0 &= m^n \int dq\,q^{2n} P^0_\epsilon(q) =: m^n[ q^{2n}]_0.
\end{align}

The probability distribution $P^0_{\epsilon}(q)\sim e^{-Nm\beta(f_\epsilon(q)-f_0(q))}$ divides into 4 parts, depending on the range of $\epsilon$. For $\epsilon\ll N^{-1/4}$  the contribution to the probability distribution of $q$ of both equations (\ref{f_1}) and (\ref{f_2}) (both with $f_{\epsilon}(q)\sim \epsilon^4q^2$) is negligible, therefore it is approximately a constant in that range for all $q\in[0,q_{EA}]$. Eq.~\eqref{beyondEA} implies that $P_\epsilon^0(q)$ has an exponentially decaying tail for $q>q_{EA}$ with $e^{-Nmc_0(q-q_{EA})^3}$. We define a function $\Theta(q-q_{EA})$ which combines both properties. Instead of this plateau in $P_\epsilon^0(q)$ there should be a $\delta$ peak at $q=q_{EA}$ which we do not see in our calculation. This is due to the fact that we have neglegted finite size corrections to the free energy which are dominating in this regime and which we will implement in the next subsection.

The two solutions we found perturbatively in Eqs.~\eqref{f_1} and \eqref{f_2} both produce a probability distribution of the form $e^{-Nm\beta c_x\epsilon^4 q^2}$ with different constants $c_x$, but hold in different ranges of $\epsilon$, depending on the relation of $\epsilon^2$ and $q$. The order of $\epsilon$ determining the crossover from one regime to the other is where $\epsilon^2$ is of the same order as $q$ and $N\epsilon^4q^2$ (in the range $N^{-1/4}\ll\epsilon\ll\epsilon_0$) is of order one such as to be the dominating part of the free energy. This leads to $\epsilon\sim N^{-1/8}$ as the crossover value. Thus we get the final result
\begin{align}
P_\epsilon(q)\propto&\begin{cases}
\Theta(q-q_{EA})& \epsilon\ll N^{-1/4}\\ 
e^{-Nmc_1\epsilon^4q^2}& N^{-1/4}\ll\epsilon\ll N^{-1/8}\\ 
e^{-Nmc_2\epsilon^4q^2}& N^{-1/8}\ll\epsilon\ll\epsilon_0\\ 
e^{-Nmf(\epsilon)q^2}&\epsilon_0 < \epsilon,
\end{cases}\label{P(q)}   
\end{align}
with $c_1=\frac{2\tau +1}{16\tau}$ and $c_2=\frac{1}{16\tau(1-2\tau)}$.

\subsection{The overlap distribution at $\epsilon=0$}
In order to account for the missing $\delta$ peak in $P_\epsilon^0(q)$ for small $\epsilon$, we have to consider the finite size corrections to the free energy, which are impossible to calculate. However, not all is lost since at least we know that at $\epsilon=0$ they grow as $N^{1-y}$ with $y=2/5$\cite{Braun:2006}, and we expect that this scaling is independent of the value of $\epsilon$. We can therefore trust our result derived above when $Nm\beta (f_\epsilon(q_{EA})-f_0)\gg N^{3/5}$. Since $q_{EA}=\mathcal{O}(1)$, it follows that this is the case when $N^{3/5}\ll N\epsilon^4$ or $N^{-1/10}\ll \epsilon$. We then have
\begin{align}
P_\epsilon(q)\propto&\begin{cases}
\delta_{FS}(q-q_{EA})& \epsilon\ll N^{-1/10}\\  e^{-Nc_2\epsilon^4q^2}& N^{-1/10}\ll\epsilon\ll\epsilon_0\\ e^{-Nf(\epsilon)q^2}&\epsilon_0 < \epsilon.
\end{cases}\label{P(q)_final}   
\end{align}
The function $\delta_{FS}(q-q_{EA})$ stands for a function which goes to a $\delta$ peak in the thermodynamic limit. Note that the new regime completely replaces the first two regimes we calculated in Eq.~(\ref{P(q)}).

\begin{figure}
\includegraphics[width=\linewidth]{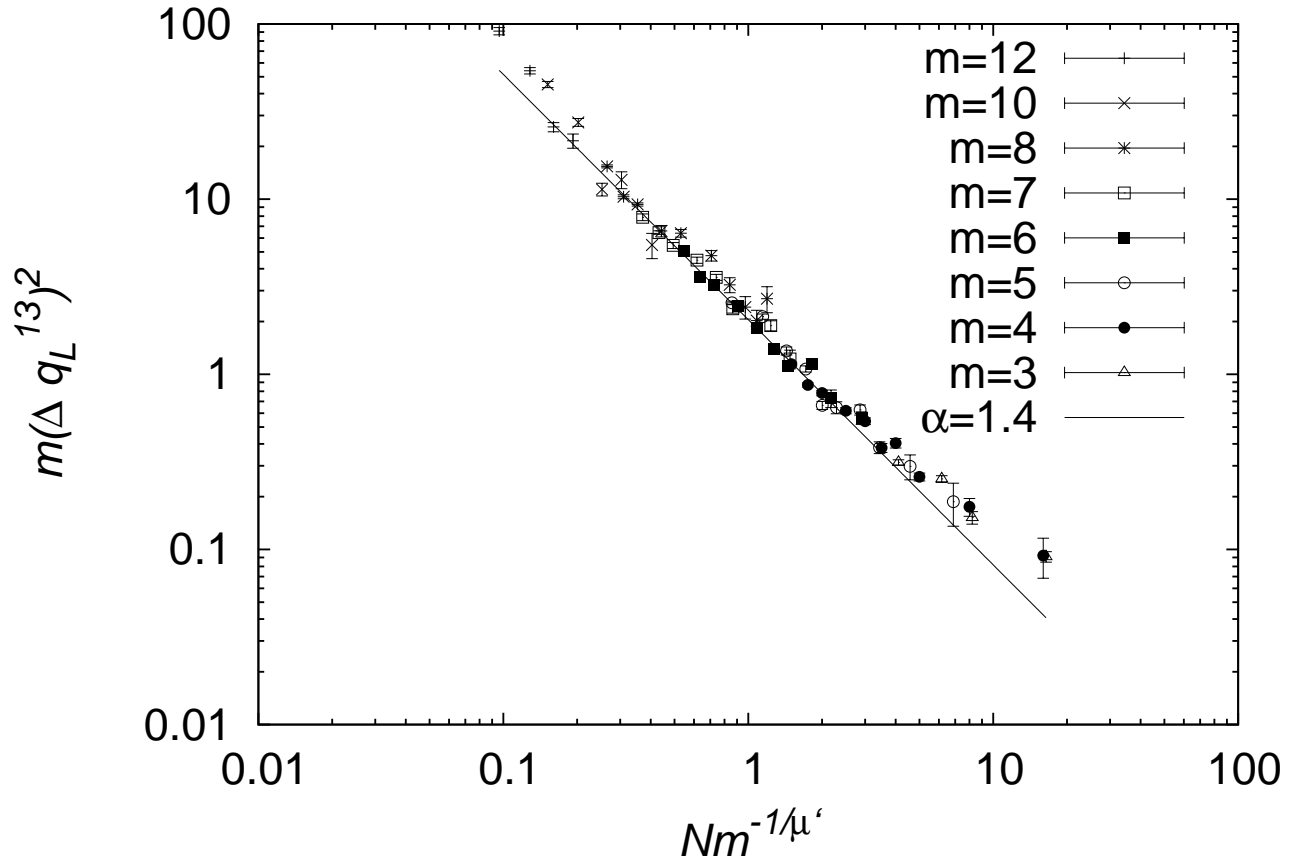}
\caption{Finite size scaling plot of the variance of the link overlap at temperature $T=0.6$. See text for explanation.}
\label{simulation}
\end{figure}
What we need in the actual calculation of the sample-to-sample
fluctuations of the free energy are the expressions $[(q_L^{13})^2] -
[q_L^{13}]^2$ and $[q_L^{13}]$. In the regimes where we have an
explicit expression for $P_\epsilon^0(q)$, we can calculate this
directly (see below). In the regime just found, however, we do not
have this information available. We must therefore resort to other
methods and use a finite size scaling ansatz of the form
\begin{align}
\frac{1}{m^2}([(q_L^{13})^2] - [q_L^{13}]^2) &= N^{-\alpha}\mathcal{F}_m(N^\beta \epsilon)
\label{scalevariance}
\end{align}
for $\epsilon\ll N^{-1/10}$ and with a scaling function $\mathcal{F}_m(x)$ whose properties will be discussed below. 
Similarly, we assume that in this regime $[q_L^{13}]$ can be written as
\begin{align}
\frac 1m [q_L^{13}] &= \mathcal{G}_m(N^\rho\epsilon)
\label{scalemean}
\end{align}
with an exponent $\rho$ and a scaling function $\mathcal{G}_m(x)$. These scaling functions and exponents will enter the calculation of the fluctuations below.

We assume $m\to\infty$ such that the system is replica symmetric. Therefore, the left hand side of Eq.~\eqref{scalevariance}, which is the variance of $q_L^{13}/m$, goes to $0$ for $N\to\infty$ since it is just the squared width of the peak in $P_\epsilon^0(q_L)$. The exponents $\alpha$ and $\beta$ are unknown, but we can obtain $\alpha$ from a simulation at $\epsilon=0$ by measuring the width of the peak in the distribution of $q_L$. This is shown in Fig.~\ref{simulation}. There is, however, a complication involved. In a simulation, $m$ must be finite, and for finite $m$, the thermodynamic limit is a replica symmetry broken phase and the variance of $q_L^{13}$ will \textit{not} tend to $0$, thus apparently making $\alpha=0$. In order to overcome this problem, we recall that in Refs.~\onlinecite{Hastings:2000,Aspelmeier:2004a} it was shown that at $T=0$ there is a critical number of spin components $n_0\sim N^{\mu'}$ (with an exponent $\mu'=2/5$) above which the system does not depend on the number of components any more and is thus identical to the replica symmetric $m=\infty$ limit. We conjecture that something similar happens at finite temperature, i.e.\ when $m\gg N^{\mu'}$, the system is in a replica symmetric phase, whereas for $m\ll N^{\mu'}$ it is in a different phase. For this reason we have plotted the variance $(\Delta q_L^{13})^2$ of $q_L^{13}$ against $x=Nm^{-1/\mu'}$, such that we can expect replica symmetric behavior for small $x$ and a crossover to a constant variance for large $x$, and this is precisely what can be observed (although the crossover is so slow that we do not see the expected plateau yet). For the determination of $\alpha$ only the data for small $x$ are relevant.

For the simulation, we have implemented a parallel tempering Monte Carlo algorithm using
system sizes from $N=64$ to $N=216$ with up to $29$ different
temperatures in the range of $[0.6:1.6]$. In that way we produced for
two different replicas (with the same set of coupling constants since we are considering $\epsilon=0$) at least $64$ statistically independent sets of spin configuration for every temperature. From these, we calculated the variance of the link
overlap $(\Delta q_L^{13})^2 = [(q_L^{13})^2]-[q_L^{13}]^2$.  According to the scaling ansatz above, $m^{-2} (\Delta q_L^{13})^2 \sim (Nm^{-1/\mu'})^{-\alpha} m^{-\alpha/\mu'}$ at $\epsilon=0$, such that $m^{\alpha/\mu'-2}(\Delta q_L^{13})^2 \sim (Nm^{-1/\mu'})^{-\alpha}$. The data in Fig.~\ref{simulation} shows on the one hand that $\alpha/\mu'-2 \approx 1$ and on the other hand that $\alpha\approx 1.4$. The observation $\alpha/\mu'-2\approx 1$ implies $\alpha\approx \frac 65$. Together with the second, more direct, observation of $\alpha$, the data shows that $\alpha\ge \frac 65$, and as we will see below, this is all we need to know.

\section{Calculating $[q^n]_0$ and the sample-to-sample fluctuations of the free energy}\label{fluctuations}
The remaining task is to calculate $[q^n]_0$ and insert it into equation (\ref{final_connection}). To do so, we use steepest descent methods to write for the regime $\epsilon$ with $N^{-1/10}\ll\epsilon\ll\epsilon_0$ (with $P_\epsilon^0(q)=\frac{1}{\mathcal{N}_q}e^{-\beta Nmf_\epsilon(q)}$ and the normalization constant $\mathcal{N}_q$)
\begin{align}
[q^n]_0&=\frac{1}{\mathcal{N}_q}\int_0^\infty dq\,q^n e^{-Nmc_x\epsilon^4q^2+\mathcal{O}(q^4)},
\end{align}
neglect the term of order $q^4$ in the exponent (note that we set the upper bound from $1$ to $\infty$, which introduces only exponentially small errors), and get (with $\mathcal{N}_q=\frac{\Gamma(\frac{1}{2})}{2\sqrt{Nmc_x\epsilon^4}}$)
\begin{align}
[q^n]_0&=\frac{1}{\Gamma(\frac{1}{2})}\int_0^\infty dx\,x^{\frac{n-1}{2}}(Nmc_x\epsilon^4)^{-n/2}e^{-x}\nonumber\\
&=\frac{(Nmc_x\epsilon^4)^{-n/2}}{\Gamma(\frac{1}{2})}\Gamma(\frac{n+1}{2}).
\end{align}
We then have
\begin{align}
[q^2]_0&=\frac{1}{2Nmc_x\epsilon^4}\label{q^2}\\
[q^4]_0&=\frac{3}{(2Nmc_x\epsilon^4)^2}\label{q^4}.
\end{align}
With this, we calculate the sample-to-sample-fluctuations through Eqs.~(\ref{final_connection}) and \eqref{final_connection2} by taking the leading order in $N$ of every integral into account. The first integrals, $I_{21}$, (with $f_2(\epsilon)=2\epsilon^3+\mathcal{O}(\epsilon^5)$) separates into three integration intervals corresponding to our three regimes. We neglect the range of $\epsilon>\epsilon_0$, because it gives a contribution of order $1$ which we are not interested in. For the part $\epsilon\ll N^{-1/10}$  we have the scaling ansatz for $[(q_L^{13})^2]-[q_L^{13}]^2= m^2 N^{-\alpha}\mathcal{F}_m(N^\beta\epsilon)$. This ansatz must match $m^2([q^4]_0 - [q^2]_0^2)$ from the neighboring regime at the crossover point $\epsilon=N^{-1/10}$, which there goes as $N^{-2}\epsilon^{-8}\sim N^{-6/5}$ (see Eqs.~\eqref{q^2} and \eqref{q^4}). This implies that the scaling function $\mathcal{F}_m(x)$ decays as $x^{-\gamma}/m^2$ for $x\to\infty$ with an exponent $\gamma$ obeying $\gamma(\beta-\frac{1}{10}) = \frac{6}{5}-\alpha$. 

This leads to
\begin{align}\label{Delta_F}
I_{21}&=\int_0^{N^{-1/10}}d\epsilon\hspace{1ex} 2\epsilon^3N^{-\alpha}\mathcal{F}_m(N^\beta\epsilon)+\int_{N^{-1/10}}^{\epsilon_0}d\epsilon\hspace{1ex} 2\epsilon^3\frac{1}{2(Nc_2\epsilon^4)^2}\nonumber\\
&\sim2N^{-\alpha-4\beta}\int_0^{N^{\beta-1/10}}dx\mathcal{F}_m(x)x^3+N^{-2}\int_{N^{-1/10}}^{\epsilon_0}d\epsilon\frac{\epsilon^{-5}}{c_2^2}+\ldots\nonumber\\
&\sim{\rm const.}_1N^{-\alpha-\frac{4}{\gamma}(\frac 65 - \alpha)-\frac{2}{5}}+{\rm const.}_2 N^{-8/5},
\end{align}
as long as $\gamma>4$. If $\gamma<4$, the first part of the integral is of order $\mathcal{O}(N^{-8/5})$.

The second integral of Eq.~(\ref{final_connection}), $I_{22}$, can be estimated in a similar way with 
the scaling ansatz
$\frac 1m [q_L^{13}]=\mathcal{G}_m(N^\rho\epsilon)$. The scaling function $\mathcal{G}_m(x)$ should have the properties
$\mathcal{G}_m(x)\stackrel{x\to\infty}{\rightarrow}x^{-\eta}$ and
$\mathcal{G}_m(x)\stackrel{x\to 0}{\rightarrow}{\rm const.}$ and should
scale as $N^{-1}\epsilon^{-4}$ for $\epsilon=N^{-\frac{1}{10}}$ in order to match the neighboring regime, which
yields the scaling relation $\rho=\frac{3}{5\eta}+\frac{1}{10}$. This gives rise to a contribution
\begin{align}
I_{22} &= \int_0^{N^{-1/10}}d\epsilon\epsilon^3\mathcal{G}_m(N^\rho\epsilon)+\int_{N^{-1/10}}^{\epsilon_0}d\epsilon\frac{\epsilon^3}{2Nc_2\epsilon^4}+\ldots \nonumber \\
&= \text{const.}_3 N^{-\frac{12}{5\eta}-\frac{2}{5}} + \text{const.}_4 \frac{\log N}{N},
\end{align} 
provided that $\eta > 4$. If $\eta<4$ the first part of the integral is  $\mathcal{O}(N^{-1})$ instead.
We therefore find the scaling exponent of the
sample-to-sample-fluctuations with the system size, $\Delta F_N\sim
N^{\mu}$, to be 
\begin{align}
\mu =\max\left[\frac{1}{5},
\frac 45 - \frac{\alpha}{2} - \frac{2}{\max(\gamma,4)}\left(\frac 65 - \alpha\right),\frac{3}{10} -\frac{6}{5\max(\eta,4)}\right]. 
\label{final_result}
\end{align}

As decribed in section \ref{connection}, we have another, slightly
different route to chaos and will use it now to check for consitency with the above
result. From Eq.~\eqref{final_connection2} we take the first integral
$I_{11} = m^2\int_0^{\epsilon_0}d\epsilon
f_1(\epsilon)([q^4]-[q^2]^2)$ and use the same scaling ansatz as
above for equation (\ref{Delta_F}) and obtain contributions of the form
\begin{align}  
I_{11} &= m^2\int_0^{\epsilon_0}d\epsilon f_1(\epsilon)([q^4]-[q^2]^2) \sim {\rm const.}_5 N^{-\alpha-\frac{3}{\gamma}(\frac 65  -\alpha)-\frac{3}{10}} + {\rm const.}_6 N^{-3/2}+\ldots
\end{align}
This integral is positive and has a negative prefactor in Eq.~\eqref{final_connection2}. The fluctuations can not be negative, however, therefore the leading order of this term \textit{must} be compensated by the second integral $I_{12}$. 
We use the scaling function $\mathcal{G}_m(N^\rho\epsilon)$ again, which yields
\begin{align}
I_{12} &= m\int_0^{\epsilon_0}d\epsilon g_1(\epsilon)[q^2] \sim {\rm const.}_7 N^{-\frac{3}{5\eta}-\frac{1}{10}}+{\rm const.}_8 N^{-7/10}.
\end{align}
This second integral, together with its leading prefactor $N$ from Eq.~\eqref{final_connection2}, must at least cancel the term of order $\mathcal{O}(N^{1/2})$ which is contained in $N^2 I_{11}$. This is only possible if $\eta\ge\frac 32$. Hence we obtain a limit on $\eta$ and no contradiction to the result derived above.

\section{Conclusion}

The purpose of this work was to calculate the finite size scaling of the sample-to-sample fluctuations of the free energy in the $m$-component vector spin glass in the limit of large $m$. The result is Eq.~\eqref{final_result}. Although this equation looks  unpromising at first sight, it is in fact very informative. First, we have the solid result $\mu\ge \frac 15$. Second, from our numerical work we know that $\alpha\ge \frac 65$. But under this condition, the second term in the $\max$ function in Eq.~\eqref{final_result} is $\le \frac 15$ and can thus simply be omitted.  Third, the last term in the $\max$ function is $\le \frac{3}{10}$, such that we obtain $\frac 15 \le \mu \le  \frac{3}{10}$. Even better, the third term is greater than $\frac 15$ only for $\eta>12$, which seems an unlikely large value. We therefore conjecture that $\mu=\frac 15$ is in fact the exact answer. But be that as it may, the exponent $\eta$ could easily be measured in simulations, and work along these lines is in progress.

Since our result is not an exact mathematical proof, we will summarize here the main assumptions on which it rests because they may have been obscured by the technicalities. The first ingredient is the connection between the fluctuations and bond chaos of the link overlap, Eqs.~\eqref{final_connection} and \eqref{final_connection2}. They are mathematically exact equalities and pose no problem. The second ingredient is the calculation of bond chaos. This is done using large deviation statistics and replica theory. We believe that replica theory in principle gives the correct results. It became apparent, however, that for the $m$ component spin glass the ``small'' deviations play a crucial role. The small deviations statistics are caused by finite size corrections of the free energy, which can not be calculated within replica theory. However, by reference to earlier results \cite{Braun:2006} we know at least the finite size scaling of the free energy corrections and can thus estimate the point where our large deviation calculation becomes valid. The region of the small deviations is then covered by a scaling ansatz, which is assumed to cross over smoothly to the region of the large deviations. The introduction of the scaling ansatz also introduced a number of unknown exponents. However, only three of these exponents are actually relevant for our results, and one of them, $\alpha$, has been measured experimentally. Moreover, the precise values of the exponents are largely irrelevant: if $\alpha\ge \frac 65$ (as we have checked numerically) and $\eta\le 12$, then $\mu=\frac 15$ is exact.

The result $\mu=\frac 15$ (or even the range $\frac 15 \le \mu \le
\frac{3}{10}$) is very interesting because it demonstrates that the
large-$m$ model is fundamentally different from the spherical
model\cite{Kosterlitz:1976}, even though their free energies are
identical\cite{Almeida:1978b}. 
While the width of the distribution of ground state energies of the
spherical spin glass scales as $N^{1/3}$, this behavior is definitely
ruled out by our results. 

It would be interesting to see whether a result similar to ours could be obtained using the methods of Ref.~\onlinecite{Parisi:2010}. We believe a corresponding replica calculation for the large-$m$ model ought to be feasible.

\acknowledgements

This work was supported by the German Science Foundation (DFG) through grant no.\ AS 136/2-1.

\bibliography{LiteraturDB,cond-mat}

\begin{thebibliography}{29}
\expandafter\ifx\csname natexlab\endcsname\relax\def\natexlab#1{#1}\fi
\expandafter\ifx\csname bibnamefont\endcsname\relax
  \def\bibnamefont#1{#1}\fi
\expandafter\ifx\csname bibfnamefont\endcsname\relax
  \def\bibfnamefont#1{#1}\fi
\expandafter\ifx\csname citenamefont\endcsname\relax
  \def\citenamefont#1{#1}\fi
\expandafter\ifx\csname url\endcsname\relax
  \def\url#1{\texttt{#1}}\fi
\expandafter\ifx\csname urlprefix\endcsname\relax\def\urlprefix{URL }\fi
\providecommand{\bibinfo}[2]{#2}
\providecommand{\eprint}[2][]{\url{#2}}

\bibitem[{\citenamefont{M\'ezard et~al.}(1987)\citenamefont{M\'ezard, Parisi,
  and Virasoro}}]{Mezard:1987}
\bibinfo{author}{\bibfnamefont{M.}~\bibnamefont{M\'ezard}},
  \bibinfo{author}{\bibfnamefont{G.}~\bibnamefont{Parisi}}, \bibnamefont{and}
  \bibinfo{author}{\bibfnamefont{M.}~\bibnamefont{Virasoro}},
  \emph{\bibinfo{title}{Spin Glass Theory and Beyond}}
  (\bibinfo{publisher}{World Scientific}, \bibinfo{address}{Singapore},
  \bibinfo{year}{1987}).

\bibitem[{\citenamefont{Parisi et~al.}(1993{\natexlab{a}})\citenamefont{Parisi,
  Ritort, and Slanina}}]{Parisi:1993}
\bibinfo{author}{\bibfnamefont{G.}~\bibnamefont{Parisi}},
  \bibinfo{author}{\bibfnamefont{F.}~\bibnamefont{Ritort}}, \bibnamefont{and}
  \bibinfo{author}{\bibfnamefont{F.}~\bibnamefont{Slanina}},
  \bibinfo{journal}{J. Phys. A} \textbf{\bibinfo{volume}{26}},
  \bibinfo{pages}{247} (\bibinfo{year}{1993}{\natexlab{a}}).

\bibitem[{\citenamefont{Parisi et~al.}(1993{\natexlab{b}})\citenamefont{Parisi,
  Ritort, and Slanina}}]{Parisi:1993a}
\bibinfo{author}{\bibfnamefont{G.}~\bibnamefont{Parisi}},
  \bibinfo{author}{\bibfnamefont{F.}~\bibnamefont{Ritort}}, \bibnamefont{and}
  \bibinfo{author}{\bibfnamefont{F.}~\bibnamefont{Slanina}},
  \bibinfo{journal}{J. Phys. A} \textbf{\bibinfo{volume}{26}},
  \bibinfo{pages}{3775} (\bibinfo{year}{1993}{\natexlab{b}}).

\bibitem[{\citenamefont{Palassini and Caracciolo}(1999)}]{Palassini:1999}
\bibinfo{author}{\bibfnamefont{M.}~\bibnamefont{Palassini}} \bibnamefont{and}
  \bibinfo{author}{\bibfnamefont{S.}~\bibnamefont{Caracciolo}},
  \bibinfo{journal}{Phys. Rev. Lett.} \textbf{\bibinfo{volume}{82}},
  \bibinfo{pages}{5128} (\bibinfo{year}{1999}).

\bibitem[{\citenamefont{Drossel et~al.}(2000)\citenamefont{Drossel, Bokil,
  Moore, and Bray}}]{Drossel:2000}
\bibinfo{author}{\bibfnamefont{B.}~\bibnamefont{Drossel}},
  \bibinfo{author}{\bibfnamefont{H.}~\bibnamefont{Bokil}},
  \bibinfo{author}{\bibfnamefont{M.~A.} \bibnamefont{Moore}}, \bibnamefont{and}
  \bibinfo{author}{\bibfnamefont{A.~J.} \bibnamefont{Bray}},
  \bibinfo{journal}{Eur. Phys. J. B} \textbf{\bibinfo{volume}{13}},
  \bibinfo{pages}{369} (\bibinfo{year}{2000}).

\bibitem[{\citenamefont{Bouchaud et~al.}(2003)\citenamefont{Bouchaud, Krzakala,
  and Martin}}]{Bouchaud:2003}
\bibinfo{author}{\bibfnamefont{J.-P.} \bibnamefont{Bouchaud}},
  \bibinfo{author}{\bibfnamefont{F.}~\bibnamefont{Krzakala}}, \bibnamefont{and}
  \bibinfo{author}{\bibfnamefont{O.~C.} \bibnamefont{Martin}},
  \bibinfo{journal}{Phys. Rev. B} \textbf{\bibinfo{volume}{68}},
  \bibinfo{pages}{224404} (\bibinfo{year}{2003}).

\bibitem[{\citenamefont{Boettcher}(2004)}]{Boettcher:2004a}
\bibinfo{author}{\bibfnamefont{S.}~\bibnamefont{Boettcher}},
  \bibinfo{journal}{Eur. Phys. J. B} \textbf{\bibinfo{volume}{38}},
  \bibinfo{pages}{83} (\bibinfo{year}{2004}).

\bibitem[{\citenamefont{Billoire}(2006)}]{Billoire:2006}
\bibinfo{author}{\bibfnamefont{A.}~\bibnamefont{Billoire}},
  \bibinfo{journal}{Phys. Rev. B} \textbf{\bibinfo{volume}{73}},
  \bibinfo{pages}{132201} (\bibinfo{year}{2006}).

\bibitem[{\citenamefont{Aspelmeier et~al.}(2008)\citenamefont{Aspelmeier,
  Billoire, Marinari, and Moore}}]{Aspelmeier:2008c}
\bibinfo{author}{\bibfnamefont{T.}~\bibnamefont{Aspelmeier}},
  \bibinfo{author}{\bibfnamefont{A.}~\bibnamefont{Billoire}},
  \bibinfo{author}{\bibfnamefont{E.}~\bibnamefont{Marinari}}, \bibnamefont{and}
  \bibinfo{author}{\bibfnamefont{M.~A.} \bibnamefont{Moore}},
  \bibinfo{journal}{J. Phys. A} \textbf{\bibinfo{volume}{41}},
  \bibinfo{pages}{324008} (\bibinfo{year}{2008}).

\bibitem[{\citenamefont{Parisi and Rizzo}(2009)}]{Parisi:2009}
\bibinfo{author}{\bibfnamefont{G.}~\bibnamefont{Parisi}} \bibnamefont{and}
  \bibinfo{author}{\bibfnamefont{T.}~\bibnamefont{Rizzo}},
  \bibinfo{journal}{Phys. Rev. B} \textbf{\bibinfo{volume}{79}},
  \bibinfo{pages}{134205} (\bibinfo{year}{2009}).

\bibitem[{\citenamefont{Parisi and Rizzo}()}]{Parisi:2010}
\bibinfo{author}{\bibfnamefont{G.}~\bibnamefont{Parisi}} \bibnamefont{and}
  \bibinfo{author}{\bibfnamefont{T.}~\bibnamefont{Rizzo}},
  \emph{\bibinfo{title}{Universality and deviations in disordered systems}},
  \bibinfo{howpublished}{arXiv:0901.1100v1 [cond-mat.dis-nn]}.

\bibitem[{\citenamefont{Sherrington and Kirkpatrick}(1975)}]{Sherrington:1975}
\bibinfo{author}{\bibfnamefont{D.}~\bibnamefont{Sherrington}} \bibnamefont{and}
  \bibinfo{author}{\bibfnamefont{S.}~\bibnamefont{Kirkpatrick}},
  \bibinfo{journal}{Phys. Rev. Lett.} \textbf{\bibinfo{volume}{35}},
  \bibinfo{pages}{1792} (\bibinfo{year}{1975}).

\bibitem[{\citenamefont{Biroli et~al.}(2007)\citenamefont{Biroli, Bouchaud, and
  Potters}}]{Biroli:2007}
\bibinfo{author}{\bibfnamefont{G.}~\bibnamefont{Biroli}},
  \bibinfo{author}{\bibfnamefont{J.-P.} \bibnamefont{Bouchaud}},
  \bibnamefont{and} \bibinfo{author}{\bibfnamefont{M.}~\bibnamefont{Potters}},
  \bibinfo{journal}{J. Stat. Mech.} \textbf{\bibinfo{volume}{2007}},
  \bibinfo{pages}{P07019} (\bibinfo{year}{2007}).

\bibitem[{\citenamefont{Tracy and Widom}(2000)}]{Tracy:2000}
\bibinfo{author}{\bibfnamefont{C.~A.} \bibnamefont{Tracy}} \bibnamefont{and}
  \bibinfo{author}{\bibfnamefont{H.}~\bibnamefont{Widom}}, in
  \emph{\bibinfo{booktitle}{Calogero-Moser-Sutherland models}}, edited by
  \bibinfo{editor}{\bibfnamefont{J.~F.} \bibnamefont{van Diejen}}
  \bibnamefont{and} \bibinfo{editor}{\bibfnamefont{L.}~\bibnamefont{Vinet}}
  (\bibinfo{publisher}{Springer Verlag}, \bibinfo{address}{New York},
  \bibinfo{year}{2000}), no.~\bibinfo{number}{4} in \bibinfo{series}{CRM Series
  in Mathematical Physics}, pp. \bibinfo{pages}{461--472}.

\bibitem[{\citenamefont{Kondor}(1983)}]{Kondor:1983b}
\bibinfo{author}{\bibfnamefont{I.}~\bibnamefont{Kondor}}, \bibinfo{journal}{J.
  Phys. A} \textbf{\bibinfo{volume}{16}}, \bibinfo{pages}{L127}
  (\bibinfo{year}{1983}).

\bibitem[{\citenamefont{Crisanti et~al.}(1992)\citenamefont{Crisanti, Paladin,
  Sommers, and Vulpiani}}]{Crisanti:1992}
\bibinfo{author}{\bibfnamefont{A.}~\bibnamefont{Crisanti}},
  \bibinfo{author}{\bibfnamefont{G.}~\bibnamefont{Paladin}},
  \bibinfo{author}{\bibfnamefont{H.-J.} \bibnamefont{Sommers}},
  \bibnamefont{and} \bibinfo{author}{\bibfnamefont{A.}~\bibnamefont{Vulpiani}},
  \bibinfo{journal}{J. Phys. I France} \textbf{\bibinfo{volume}{2}},
  \bibinfo{pages}{1325} (\bibinfo{year}{1992}).

\bibitem[{\citenamefont{Parisi and Rizzo}(2010)}]{Parisi:2010a}
\bibinfo{author}{\bibfnamefont{G.}~\bibnamefont{Parisi}} \bibnamefont{and}
  \bibinfo{author}{\bibfnamefont{T.}~\bibnamefont{Rizzo}}, \bibinfo{journal}{J.
  Phys. A: Math. Gen.} \textbf{\bibinfo{volume}{43}} (\bibinfo{year}{2010}).

\bibitem[{\citenamefont{Aspelmeier}(2008{\natexlab{a}})}]{Aspelmeier:2008}
\bibinfo{author}{\bibfnamefont{T.}~\bibnamefont{Aspelmeier}},
  \bibinfo{journal}{Phys. Rev. Lett.} \textbf{\bibinfo{volume}{100}},
  \bibinfo{pages}{117205} (\bibinfo{year}{2008}{\natexlab{a}}).

\bibitem[{\citenamefont{Aspelmeier}(2008{\natexlab{b}})}]{Aspelmeier:2008a}
\bibinfo{author}{\bibfnamefont{T.}~\bibnamefont{Aspelmeier}},
  \bibinfo{journal}{J. Stat. Mech.} p. \bibinfo{pages}{P04018}
  (\bibinfo{year}{2008}{\natexlab{b}}).

\bibitem[{\citenamefont{Kosterlitz et~al.}(1976)\citenamefont{Kosterlitz,
  Thouless, and Jones}}]{Kosterlitz:1976}
\bibinfo{author}{\bibfnamefont{J.~M.} \bibnamefont{Kosterlitz}},
  \bibinfo{author}{\bibfnamefont{D.~J.} \bibnamefont{Thouless}},
  \bibnamefont{and} \bibinfo{author}{\bibfnamefont{R.~C.} \bibnamefont{Jones}},
  \bibinfo{journal}{Phys. Rev. Lett.} \textbf{\bibinfo{volume}{36}},
  \bibinfo{pages}{1217} (\bibinfo{year}{1976}).

\bibitem[{\citenamefont{Andreanov et~al.}(2004)\citenamefont{Andreanov,
  Barbieri, and Martin}}]{Andreanov:2004}
\bibinfo{author}{\bibfnamefont{A.}~\bibnamefont{Andreanov}},
  \bibinfo{author}{\bibfnamefont{F.}~\bibnamefont{Barbieri}}, \bibnamefont{and}
  \bibinfo{author}{\bibfnamefont{O.~C.} \bibnamefont{Martin}},
  \bibinfo{journal}{Eur. Phys. J. B} \textbf{\bibinfo{volume}{41}},
  \bibinfo{pages}{365} (\bibinfo{year}{2004}).

\bibitem[{\citenamefont{de~Almeida et~al.}(1978)\citenamefont{de~Almeida,
  Jones, Kosterlitz, and Thouless}}]{Almeida:1978b}
\bibinfo{author}{\bibfnamefont{J.~R.~L.} \bibnamefont{de~Almeida}},
  \bibinfo{author}{\bibfnamefont{R.~C.} \bibnamefont{Jones}},
  \bibinfo{author}{\bibfnamefont{J.~M.} \bibnamefont{Kosterlitz}},
  \bibnamefont{and} \bibinfo{author}{\bibfnamefont{D.~J.}
  \bibnamefont{Thouless}}, \bibinfo{journal}{J. Phys. C}
  \textbf{\bibinfo{volume}{11}}, \bibinfo{pages}{L871} (\bibinfo{year}{1978}).

\bibitem[{\citenamefont{Viana}(1988)}]{Viana:1988}
\bibinfo{author}{\bibfnamefont{L.}~\bibnamefont{Viana}}, \bibinfo{journal}{J.
  Phys. A} \textbf{\bibinfo{volume}{21}}, \bibinfo{pages}{803}
  (\bibinfo{year}{1988}).

\bibitem[{\citenamefont{Aspelmeier}(2008{\natexlab{c}})}]{Aspelmeier:2008b}
\bibinfo{author}{\bibfnamefont{T.}~\bibnamefont{Aspelmeier}},
  \bibinfo{journal}{J. Phys. A} \textbf{\bibinfo{volume}{41}},
  \bibinfo{pages}{205005} (\bibinfo{year}{2008}{\natexlab{c}}).

\bibitem[{\citenamefont{Hastings}(2000)}]{Hastings:2000}
\bibinfo{author}{\bibfnamefont{M.~B.} \bibnamefont{Hastings}},
  \bibinfo{journal}{J. Stat. Phys.} \textbf{\bibinfo{volume}{99}},
  \bibinfo{pages}{171} (\bibinfo{year}{2000}).

\bibitem[{\citenamefont{Aspelmeier and Moore}(2004)}]{Aspelmeier:2004a}
\bibinfo{author}{\bibfnamefont{T.}~\bibnamefont{Aspelmeier}} \bibnamefont{and}
  \bibinfo{author}{\bibfnamefont{M.~A.} \bibnamefont{Moore}},
  \bibinfo{journal}{Phys. Rev. Lett.} \textbf{\bibinfo{volume}{92}},
  \bibinfo{pages}{077201} (\bibinfo{year}{2004}).

\bibitem[{\citenamefont{Lee et~al.}(2005)\citenamefont{Lee, Dhar, and
  Young}}]{Lee:2005a}
\bibinfo{author}{\bibfnamefont{L.~W.} \bibnamefont{Lee}},
  \bibinfo{author}{\bibfnamefont{A.}~\bibnamefont{Dhar}}, \bibnamefont{and}
  \bibinfo{author}{\bibfnamefont{A.~P.} \bibnamefont{Young}},
  \bibinfo{journal}{Phys. Rev. E} \textbf{\bibinfo{volume}{71}},
  \bibinfo{pages}{036146} (\bibinfo{year}{2005}).

\bibitem[{\citenamefont{Braun and Aspelmeier}(2006)}]{Braun:2006}
\bibinfo{author}{\bibfnamefont{A.}~\bibnamefont{Braun}} \bibnamefont{and}
  \bibinfo{author}{\bibfnamefont{T.}~\bibnamefont{Aspelmeier}},
  \bibinfo{journal}{Phys. Rev. B} \textbf{\bibinfo{volume}{74}},
  \bibinfo{pages}{144205} (\bibinfo{year}{2006}).

\bibitem[{\citenamefont{Guerra and Toninelli}(2002)}]{Guerra:2002}
\bibinfo{author}{\bibfnamefont{F.}~\bibnamefont{Guerra}} \bibnamefont{and}
  \bibinfo{author}{\bibfnamefont{F.~L.} \bibnamefont{Toninelli}},
  \bibinfo{journal}{Commun. Math. Phys.} \textbf{\bibinfo{volume}{230}},
  \bibinfo{pages}{71} (\bibinfo{year}{2002}).

\end{thebibliography}
\end{document}